\documentclass{article}
\usepackage{amsmath,amssymb,amsfonts,amsthm}
\usepackage[russian]{babel}
\usepackage{color}
\usepackage{graphicx}
\usepackage{wrapfig}

\begin{document}

\flushbottom

\renewcommand{\figurename}{Fig.}
\def\refname{References}
\def\proofname{Proof}

\newtheorem{theorem}{Theorem}
\newtheorem{proposition}{Proposition}
\newtheorem{remark}{Remark}
\newtheorem{corollary}{Corollary}

\def\tens#1{\ensuremath{\mathsf{#1}}}

\if@mathematic
   \def\vec#1{\ensuremath{\mathchoice
                     {\mbox{\boldmath$\displaystyle\mathbf{#1}$}}
                     {\mbox{\boldmath$\textstyle\mathbf{#1}$}}
                     {\mbox{\boldmath$\scriptstyle\mathbf{#1}$}}
                     {\mbox{\boldmath$\scriptscriptstyle\mathbf{#1}$}}}}
\else
   \def\vec#1{\ensuremath{\mathchoice
                     {\mbox{\boldmath$\displaystyle\mathbf{#1}$}}
                     {\mbox{\boldmath$\textstyle\mathbf{#1}$}}
                     {\mbox{\boldmath$\scriptstyle\mathbf{#1}$}}
                     {\mbox{\boldmath$\scriptscriptstyle\mathbf{#1}$}}}}
\fi

\begin{center}
{\Large\bf The phenomenon of reversal
in~the~Euler\,--\,Poincar\'{e}\,--\,Suslov nonholonomic~systems\\}

\bigskip

\footnotetext{This work was supported by the grant of the Russian
Scientific Foundation (project 14-50-00005).}

{\large\bf Valery~V.\,Kozlov\\}
\end{center}

\begin{quote}
\begin{small}
\noindent
Steklov Mathematical Institute, Russian Academy of Sciences\\
Gubkina st. 8, Moscow, 119991, Russia
\end{small}

\bigskip
\bigskip

\begin{small}
\textbf{Abstract.} In this paper, the dynamics of nonholonomic systems on
Lie groups with a~left-invariant kinetic energy and left-invariant
constraints are considered. Equations of motion form a~closed system of
differential equations on the corresponding Lie algebra. In addition, the
effect of change in the stability of steady motions of these systems with
the direction of motion reversed (the reversal found in rattleback
dynamics) is discussed. As an illustration, the~rotation of a~rigid body
with a~fixed point and the Suslov nonholonomic constraint as well as the
motion of the Chaplygin sleigh is considered.

\smallskip


\smallskip

\end{small}
\end{quote}

\clearpage

\section{Euler\,--\,Poincar\'{e}\,--\,Suslov nonholonomic systems}\label{kozlov_sec1}

Let the configuration space of a~mechanical system with $n$ degrees of
freedom represent a~Lie group $G$, and let $w_1,\ldots,w_n$ be independent
left-invariant vector fields on~$G$ (they are invariant under all left translations on
$G$). The commutators of these fields have the form
\begin{equation}
\label{kozlov_eq1.1}
[w_i,w_j]=\sum\limits_{k=1}^n c_{ij}^k w_k,
\end{equation}
where $c_{ij}^k=-c_{ji}^k$~are the structure constants of~$G$.

Let $x=(x_1,\ldots,x_n)$~be local (generalized) coordinates on
$G$ and $f(x)$~be a~smooth function. Then
$$
\dot f=\frac{\partial f}{\partial x}\cdot \dot x=\sum\limits_{i=1}^n w_i(f)\omega_i,
$$
where $w_i(f)=\frac{\partial f}{\partial x}\cdot  w_i$~is the derivative $f$ along the vector field $w_i$.
The variables $\omega$ (called quasi-velocities)
linearly depend on $\dot x$:
\begin{equation}
\label{kozlov_eq1.2}
\dot x_k=\sum\limits_{i=1}^n w_i(x_k)\omega_i,\quad 1\le k\le n.
\end{equation}
The quasi-velocities~are Cartesian coordinates on the Lie algebra~$g$
of~$G$.

Assume that the Lagrangian of the system of interest is reduced to the
kinetic energy $T$ which generates a left-invariant Riemannian metric
$(\cdot,\cdot)$ on~$G$. In this case
$$
T=\frac12\, (\dot x,\dot x)=\frac12 \Big(\sum w_i\omega_i, \sum w_j\omega_j\Big)=
    \frac12 \sum I_{ij}\omega_i\omega_j,
$$
where
$$
I_{ij}=(w_i,w_j)=\text{const},
$$
since all the vector fields $w_1,\ldots,w_n$ are left-invariant. The scalar-product
matrix $\|I_{ij}\|$ is positive definite. This is the inertia tensor of
the mechanical system.

According to Poincar\'{e}~\cite{kozlov_1} (see also \cite{kozlov_2}), the Lagrange
equations take the following form:
\begin{equation}
\label{kozlov_eq1.3}
\dot m_i=\sum\limits_{j,k=1}^n c_{ji}^k m_k\omega_j,\quad 1\le i\le n.
\end{equation}
Here
$$
m_i=\frac{\partial T}{\partial \omega_i}=\sum I_{ij}\omega_j
$$
are the components of the angular momentum of the system; this is a~vector
from the linear space~$g^*$ dual to~$g$. Eqs.~\eqref{kozlov_eq1.3} are
a~closed system of differential equations with quadratic right-hand sides
on~$g$ (or on~$g^*$). The system is often called
\textit{Euler\,--\,Poincar\'{e} equations} on a~Lie algebra. For the group
of rotations of the three-dimensional Euclidean space $SO(3)$,
Eqs.~\eqref{kozlov_eq1.3} coincide with the Euler equations from rigid
body dynamics. To completely describe the motion, one needs to add the
kinematic equations~\eqref{kozlov_eq1.2} to Eqs.~\eqref{kozlov_eq1.3}.

Now we increase the complexity of the problem by adding some constraints that are linear in the velocities and generally nonintegrable:
$$
f_1(\omega,x)=\ldots=f_m(\omega, x)=0 \quad (m<n).
$$
Of particular interest is the case
where, apart from the kinetic energy~$T$, the functions of the constraints
$f_1,\ldots,f_m$ are also left-invariant (i.\,e., explicitly independent
of~$x$). Under this condition the nonholonomic Lagrangian equations with
the constraints
\begin{equation}
\label{kozlov_eq1.4}
\dot m_i=\sum\limits_{j,k=1}^n c_{ji}^k m_k\omega_j+\sum\limits_{p=1}^m \lambda_p \frac{\partial f_p}{\partial \omega_i},\quad
f_1(\omega)=\ldots=f_m(\omega)=0
\end{equation}
form a~closed system of differential equations on the Lie algebra~$g$.

For the case $G=SO(3)$ systems with left-invariant constraints were first
studied by Suslov~\cite{kozlov_3}. He considered the rotation of a~rigid
body about a~fixed point with the following nonholonomic constraint: the
projection of the angular velocity vector $\omega\in SO(3)=\mathbb{R}^3$ onto
a~body-fixed axis equals zero. General nonholonomic systems on Lie groups
with a~left-invariant kinetic energy and left-invariant constraints were
studied in~\cite{kozlov_4,kozlov_5}. These systems are called
\textit{Euler\,--\,Poincar\'{e}\,--\,Suslov} (EPS) systems. Since the
constraints are linear in the velocity, Eqs.~\eqref{kozlov_eq1.4} admit
the energy integral
\begin{equation}
\label{kozlov_eq1.5}
T=\frac12 \sum I_{ij}\omega_i\omega_j=h=\text{const}.
\end{equation}

In Suslov's problem Eqs.~\eqref{kozlov_eq1.4} have the form
\begin{equation}
\label{kozlov_eq1.6}
I\dot \omega+\omega\times I\omega=\lambda a,\quad (a,\omega)=0.
\end{equation}
Here $a\ne 0$ is some constant vector in a~moving space. If~$a$ is an
eigenvector of the inertia operator ($Ia=\rho a$,
$\rho\in\mathbb{R}$), then it follows from~\eqref{kozlov_eq1.6} that $\omega=\text{const}$
(see~\cite{kozlov_3}). In particular, all rotations of the rigid body are stable.

In a~typical case (where this condition does not hold), each solution
$t\,{\mapsto}\,
\omega(t)$ possesses the following property:
\begin{equation}
\label{kozlov_eq1.7}
\lim\limits_{t\to \pm\infty} \omega(t)=\pm \omega_0.
\end{equation}
The motion of a~rigid body is an asymptotic transition from steady
rotation about some body-fixed axis to steady rotation with
the same angular velocity about the same axis but in the opposite
direction. We emphasize that in fixed space, as $t \to +\infty$ and $t \to -\infty$, these
rotations occur about different axes.
The angular velocities of the steady rotations~\eqref{kozlov_eq1.7} are found from
the system~\eqref{kozlov_eq1.6}. On the one hand, we have the constraint equation
$(a,\omega_0)=0$. On the other hand, performing scalar multiplication of the first
equation~\eqref{kozlov_eq1.6} by $I\omega^0$, we obtain one more equation:
$(Ia,\omega_0)=0$. This yields all steady-state solutions of the
system~\eqref{kozlov_eq1.6}:
\begin{equation}
\label{kozlov_eq1.8}
\omega_0=\mu a\times Ia,\quad \mu\in \mathbb{R}.
\end{equation}

\begin{figure}[!ht]
\centering
\includegraphics[scale=0.75]{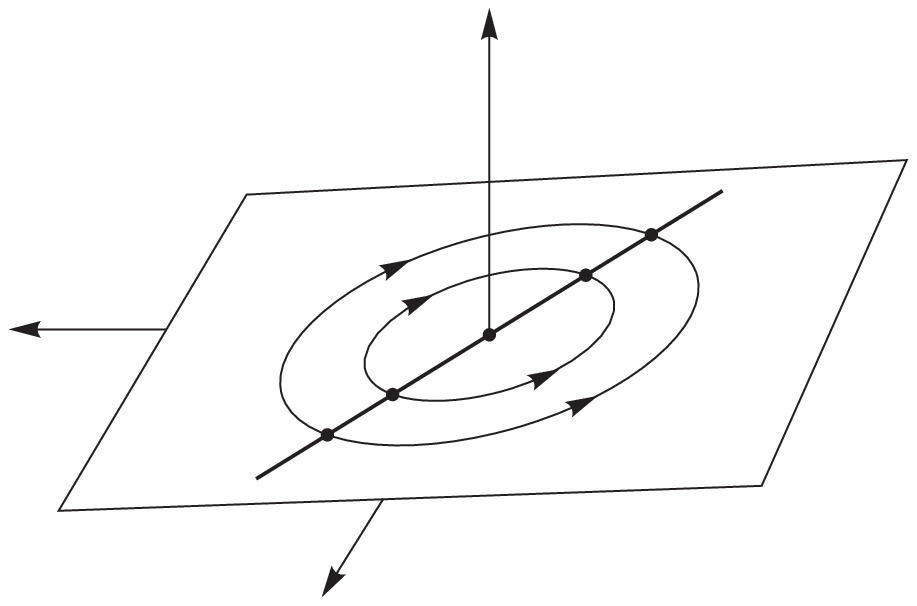}
\caption{}\label{kozlov_fig1}
\end{figure}

The phase portrait of the system~\eqref{kozlov_eq1.6} is shown in Fig.~\ref{kozlov_fig1}.

It is evident from this phase portrait that the phase flow of this system
does not admit an invariant measure with a~continuous (and even summable)
density. The conditions for the existence of an invariant measure of the
EPS equations \eqref{kozlov_eq1.4} in the multidimensional case are given
in~\cite{kozlov_4}. Equality~\eqref{kozlov_eq1.7} holds also for Suslov's problem in
the $n$-dimensional space where $G=SO(n)$ \cite{kozlov_5} (see
also~\cite{kozlov_gizs13,kozlov_gizs14}).

The absence of an invariant measure prevents one from applying
the\linebreak geometrical version of the Euler\,--\,Jacobi theorem which
was formulated in~\mbox{\cite{kozlov_gizs3,kozlov_bbm_gamilt}}, although
explicit quadratures can be obtained (see~\cite{kozlov_3}). This motivated
the development of an approach~\cite{kozlov_gizs} based on the
Hamiltonization of this system in open regions of the phase space.
In~\cite{kozlov_gizs18}, it is shown that under certain conditions the
complete system in Suslov's problem (including the evolution of the Euler
angles) can admit an additional algebraic integral of arbitrarily high
degree in velocities. The explicit form of these integrals is found
in~\cite{kozlov_gizs}, where it is also noted that in this case the angle
(in a fixed coordinate system) between the axes of limiting steady
rotations is $\pm\pi$. In the general case, when the explicit form of
first integrals is unknown, this issue is discussed in detail in
\cite{Fedoo}, and the angle between the same axes is determined by some
formula depending on the system parameters.

Using explicit formulas to solve Eqs.~\eqref{kozlov_eq1.6}
\cite{kozlov_3}, one can conclude that steady rotations
of~\eqref{kozlov_eq1.8} are unstable for~${\mu>0}$, and conversely, they
are stable for~${\mu<0}$ (they are even asymptotically stable if the
value of the total energy $h>0$ from~\eqref{kozlov_eq1.5} is fixed).

Thus, the nonholonomic Suslov top can be spun about the
axis~\eqref{kozlov_eq1.8} only in one direction: when spun in the opposite
direction, the top loses stability and, with the course of time,
begins to rotate in the opposite direction. This
property of the Suslov top, usually called {\it reversal}, makes it similar to rattlebacks, which
also exhibit the asymmetry of stability when the spin direction is
reversed \cite{kozlov_6}, see also~\cite{kozlov_7}. We note that in \cite{Kaz} the property of reversal was noticed for the Chaplygin top (a
dynamically asymmetric ball with a displaced center of mass rolling on a horizontal plane in a gravitational field).

For recent developments in the study of rattleback dynamics, see~{\bf
\cite{kozlov_gizs30,kozlov_bkzh, GG}}, where, in particular, it is shown
that the absence of an invariant measure can lead to the existence of
a~system of strange attractors in phase space.

The purpose of this note is to show the universality of the phenomenon of
stability changes when the direction of the velocity of steady motions of nonholonomic EPS systems is reversed. Note that in the general case
the absence or presence of tensor
invariants~\cite{kozlov_gizs3,kozlov_kozlov_lee,kozlov_laws} in
nonholonomic systems leads to a~kind of hierarchy of dynamical behavior
described in~\cite{kozlov_hierachy, Bizz}.

\section{The Chaplygin sleigh as an EPS system}\label{kozlov_sec2}

The Chaplygin sleigh is a~rigid body with a~nonholonomic constraint moving
on a~horizontal plane: the velocity of some point of it is always
orthogonal to the body-fixed direction (see~\cite{kozlov_8,kozlov_9}). For instance, one
can assume that a~vertical wheel which is unable to move in the direction
orthogonal to its plane is rigidly attached to the body. We show that this
nonholonomic system is an EPS system on the group of motions $E(2)$ of the
Euclidean plane.

\begin{figure}[!b]
\centering
\includegraphics{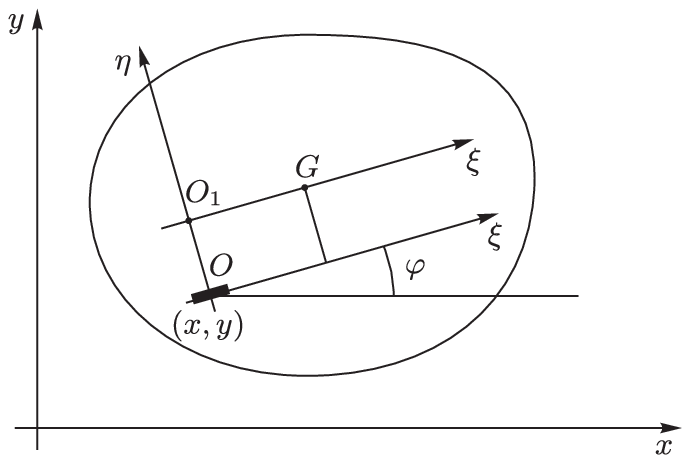}
\caption{}\label{kozlov_fig2}
\end{figure}

Without loss of generality, the rigid body itself may be assumed to be
flat. The positions of this body on the plane are determined by three
parameters: the Cartesian coordinates $x$ and $y$ of a distinguished point~$O$,
and the rotation angle $\varphi$ (Fig.~\ref{kozlov_fig2}). The independent
left-invariant fields on $E(2)$ are specified by the differential
operators
$$
X_{\xi}=\cos\varphi\,\frac{\partial }{\partial x}+\sin \varphi\,\frac{\partial }{\partial y},\quad
X_{\eta}=-\sin\varphi\,\frac{\partial }{\partial x}+\cos\varphi\,\frac{\partial }{\partial y},\quad
X_{\varphi}=\frac{\partial }{\partial \varphi}.
$$
The commutation relations~\eqref{kozlov_eq1.1} are
\begin{equation}
\label{kozlov_eq2.1}
[X_{\xi},X_{\varphi}]=-X_{\eta}, \quad
[X_{\eta},X_{\varphi}]=X_{\xi}, \quad
[X_{\xi},X_{\eta}]=0.
\end{equation}

Let $\xi$, $\eta$~be the coordinates of the center of mass of the
body~$G$ in a~moving reference frame, $u$, $v$~be the projections of the velocity of
the origin of the moving frame onto these axes, and $\omega$~be the angular velocity.
The nonintegrable constraint equation is
\begin{equation}
\label{kozlov_eq2.2}
f=v=0.
\end{equation}
This constraint is left-invariant.

Let~$m$ be the mass of the body and~$J_0$ be its inertia moment relative
to the center of mass. The kinetic energy
\begin{equation}
\label{kozlov_eq2.3}
T=\frac{m}2 \Big((u-\omega\eta)^2+(v+\omega\xi)^2\Big)+\frac{J_0\omega^2}{2}
\end{equation}
is also left-invariant: like the constraint function, it is
independent of the generalized coordinates. Hence, the Chaplygin sleigh is an
example of an EPS system on the group~$E(2)$.

Taking into account~\eqref{kozlov_eq2.1}, the EPS equations for the
Chaplygin sleigh take the form
\begin{equation}
\label{kozlov_eq2.4}
\bigg(\frac{\partial T}{\partial u}\bigg)^{\boldsymbol\cdot}=\omega\frac{\partial T}{\partial v},\quad
\bigg(\frac{\partial T}{\partial v}\bigg)^{\boldsymbol\cdot}=-\omega\frac{\partial T}{\partial u}+\lambda,\quad
\bigg(\frac{\partial T}{\partial \omega}\bigg)^{\boldsymbol\cdot}=v\frac{\partial T}{\partial u}-u\frac{\partial T}{\partial v}.
\end{equation}
Of course, one needs to add to them the constraint equation~\eqref{kozlov_eq2.2}.

The second equation serves as the basis for finding the Lagrange
multiplier (the constraint reaction), and the first and the third
equations (taking into account the constraint) have the following explicit form:
\begin{equation}
\label{kozlov_eq2.5}
\dot u-\eta\dot\omega=\xi \omega^2,\quad
(J_0+m\xi^2+m\eta^2)\dot \omega-m\eta\dot u=-m\xi u\omega.
\end{equation}
These equations should be supplemented with those describing the law of
motion of the point~$O$ and the rotation angle~$\varphi$:
\begin{equation}
\label{kozlov_eq2.5_12}
\dot x=u\cos\varphi,\quad
\dot y=u\sin\varphi,\quad
\dot \varphi=\omega.
\end{equation}
A qualitative analysis of motion of the Chaplygin sleigh for
$\eta=0$ can be found in~\cite{kozlov_9}. We supplement this analysis with some remarks.

\begin{proposition}\!\!\cite{kozlov_bm1}.
If $\eta\ne0$, then the equations of motion in the coordinate frame $O_1\xi\eta$
with the origin~$O_1$ lying
at the intersection of the straight line $O_1G$ which is parallel to
the blade and passes through the center of mass, and the straight line~$OO_1$
which is perpendicular to~$O_1G$ and passes through the contact
point of the blade, coincide with the equations of motion~\eqref{kozlov_eq2.5},
\eqref{kozlov_eq2.5_12} in the frame $O\xi\eta$ when $\eta=0$.
\end{proposition}

Thus, if we consider the trajectory of point~$O_1$ instead of the
trajectory of the contact point of the blade, then without loss of
generality we may set $\eta=0$ in the equations of
motion~\eqref{kozlov_eq2.5}.

If $\xi=0$, all motions are steady. If $\xi\ne 0$, the phase
portrait of the system~\eqref{kozlov_eq2.5} is similar in appearance to the phase portrait of
Suslov's problem (Fig.~\ref{kozlov_fig1}). In particular, the nonlinear
equations~\eqref{kozlov_eq2.5} do not admit an invariant measure with a~summable
density. The steady--state solutions of~\eqref{kozlov_eq2.5}
$$
u=u_0,\quad \omega=0
$$
correspond to the rectilinear motions of the Chaplygin sleigh. They are
stable for $\xi u_0>0$ and unstable for $\xi u_0<0$.  Thus,
during a~stable motion in a~straight line, the center of mass of the body ``outstrips''
the contact point of the wheel. Here a change in stability
occurs when the direction of motion is reversed.\looseness=1

In fact, the system's global evolution on the time interval $t\in(-\infty,+\infty)$ in this case
may be considered as a~scattering process, that is, as $t\to-\infty$, the sleigh ``starts''
its motion from some unstable steady--state solution (the
center of mass ``lags behind'' the contact point), undergoes some
evolution, and tends to some stable steady-state solution as $t\to+\infty$. At
the same time, the complete rotation angle of the axis $O\xi$ (or the axis
$O\xi_1$) is independent of the initial conditions and, according
to~\cite{kozlov_bm1}, is defined by
$$
\Delta \varphi\big|_{-\infty}^{+\infty}=\pi A,\quad
A^2=\bigg(1+\frac{J_0}{m\xi^2}\bigg).
$$

The Hamiltonian representation of the system~\eqref{kozlov_eq2.5}-\eqref{kozlov_eq2.5_12} and the redundant set of first integrals are also given
in~\cite{kozlov_bm1}.

\begin{figure}[!ht]
\centering
\includegraphics{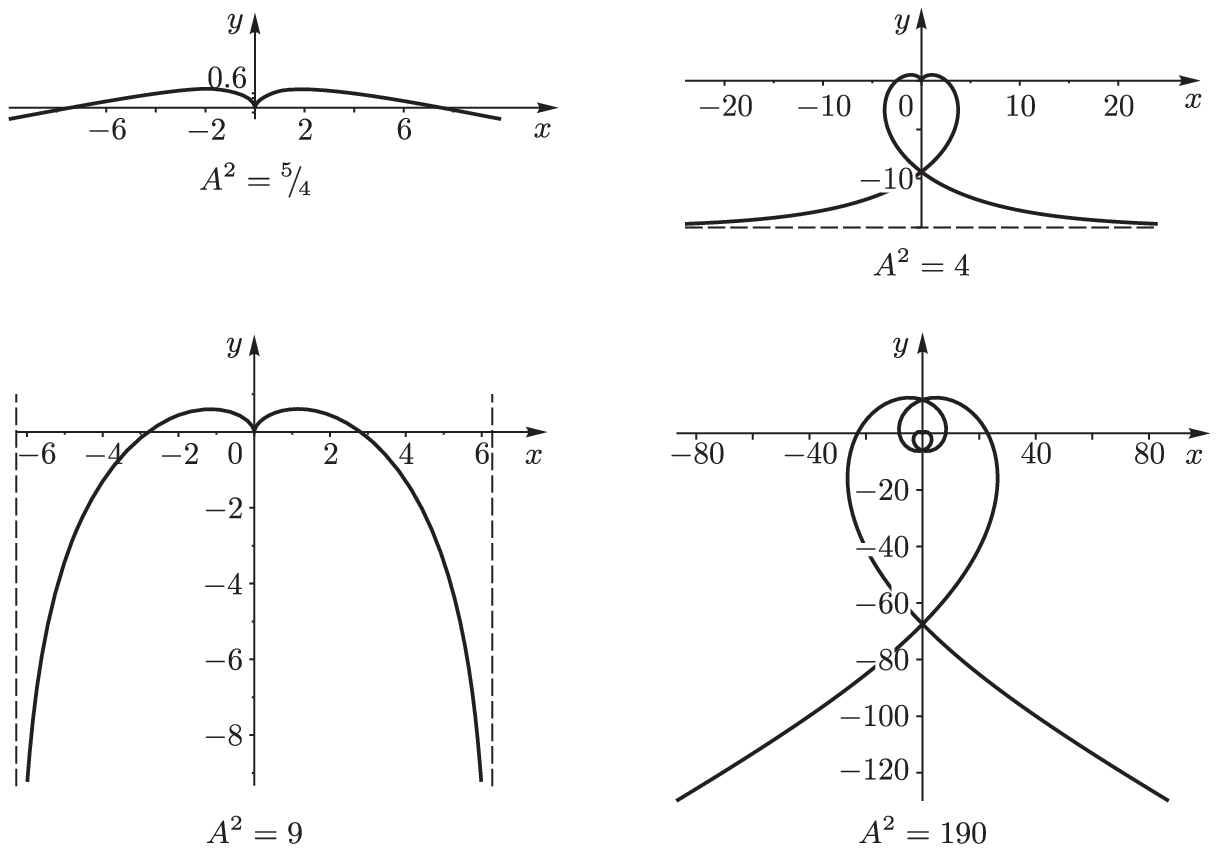}
\caption{}\label{kozlov_fig3}
\end{figure}

Typical trajectories of point~$O_1$ for various values of~$A$
are shown in Fig.~\ref{kozlov_fig3}.

Examples of other nonholonomic systems represented in Hamiltonian and
conformally Hamiltonian forms are given in~\cite{kozlov_laws,kozlov_hierachy,kozlov_bm_rolling,kozlov_bfm}. We also note that the motion of the Chaplygin sleigh
on a rotating plane is considered in \cite{SS}.

\begin{remark}
There is a~simple and natural connection between Suslov's and Chaplygin's
problems. We consider the generalized Chaplygin sleigh on a~two-dimensional sphere: this is a~spherical ``cap'' sliding on the
sphere with the nonholonomic constraint. Its dynamics are described by Suslov's equations. In the extreme case, as the radius of the sphere
increases to infinity, we obtain the Chaplygin sleigh. This
technique includes the well-known retraction of the group~$SO(3)$ to the
group~$E(3)$.
\end{remark}

\section{Anisotropic friction}\label{kozlov_sec3}

A nonholonomic model describes the dynamics of systems with ideal
constraints. There exist more realistic models which take into account
large forces of viscous friction with anisotropic Rayleigh's function,
see, for example,~\cite{kozlov_gizs26}. The\linebreak corresponding
passage to the limit with regard to the Chaplygin sleigh was studied
in~\cite{kozlov_10,kozlov_9}.

Following Carath\'{e}odory, we consider the motion of a~rigid body
on\linebreak a~horizontal plane taking into account the force of viscous
friction applied to a~fixed point of the rigid body and orthogonal to some
body-fixed direction. In the notation of Section~\ref{kozlov_sec2} the
equations of motion are described by the following closed system on the
algebra~$e(3)$:
\begin{equation}
\label{kozlov_eq3.1}
\bigg(\frac{\partial T}{\partial u}\bigg)^{\boldsymbol\cdot}=\omega\frac{\partial T}{\partial v},\quad
\bigg(\frac{\partial T}{\partial v}\bigg)^{\boldsymbol\cdot}=-\omega\frac{\partial T}{\partial u}-kv,\quad
\bigg(\frac{\partial T}{\partial \omega}\bigg)^{\boldsymbol\cdot}=v\frac{\partial T}{\partial u}-u\frac{\partial T}{\partial v}.
\end{equation}
Here, $k>0$~is the coefficient of viscous friction. In contrast
to~\eqref{kozlov_eq2.4}, there are no constraints here. We write the explicit form
of these nonlinear equations:
\begin{equation}
\label{kozlov_eq3.2}
\begin{gathered}
\dot u-\eta\dot \omega=\omega v+\xi\omega^2,\quad
\dot v+\xi \dot \omega=-\omega u+\eta\omega^2-\frac{k}{m}\,v,\\
\frac{J}{m}\,\dot \omega+\xi \dot v-\eta \dot u=-\omega(\xi u+\eta v).
\end{gathered}
\end{equation}
Here,~$J=J_0+m(\xi^2+\eta^2)$ is the moment of inertia of the body about the
distinguished point.

Like~\eqref{kozlov_eq2.4}, the system~\eqref{kozlov_eq3.2} has the family of
steady-state solutions
\begin{equation}
\label{kozlov_eq3.3}
u=u_0,\quad v=0,\quad \omega=0
\end{equation}
corresponding to the motion of a~rigid body in a~straight line with a~constant velocity. In view of the equality $\dot T=-k v^2$, the energy
does not dissipate on these solutions. In order to study the stability of
the solutions~\eqref{kozlov_eq3.3}, we linearize in their
neighborhood Eqs.~~\eqref{kozlov_eq3.2}. Omitting simple calculations, we give the characteristic
polynomial of the linearized system:
\begin{equation}
\label{kozlov_eq3.4}
f(\lambda)=\lambda\bigg[\lambda^2\frac{J_0}{m}+\lambda\frac{J_0+m\xi^2}{m^2}+\frac{\xi u_0k}{m}\bigg].
\end{equation}
The appearance of the zero root is due to the fact that the steady-state
solutions \eqref{kozlov_eq3.3} are nonisolated.

If~$\xi u_0<0$ (the contact point ``outstrips'' the center of mass when the body is moving),
the characteristic polynomial has a~real positive
root and therefore the steady motion in a~straight line is unstable
(as in the case of the Chaplygin sleigh).

We show that given $\xi u_0>0$ (when the center of mass
``outstrips'' the contact point) the steady motion is stable.
Moreover, perturbed motions asymptotically tend to one of the steady
states~\eqref{kozlov_eq3.3}. By Lyapunov's classical theorem
\cite[Section~32]{kozlov_11}, it is sufficient to show that the nonzero roots of the
characteristic polynomial~\eqref{kozlov_eq3.4} have negative real parts. But this,
in turn, follows from the positivity property of the coefficients of
the polynomial~\eqref{kozlov_eq3.4}.

We also present the asymptotic form of the roots of the
characteristic\linebreak polynomial as $k\,{\to}\,\infty$:
$$
\lambda_1=0,\quad
\lambda_2=-\frac{m\xi u_0}{J_0+m\xi^2}+O\bigg(\frac1k\bigg),\quad
\lambda_3=-\frac{J_0+m\xi^2}{J_0m}\,k+O(1).
$$
It is clear that $\lambda_3\to-\infty$ as $k\to \infty$, and
\begin{equation}
\label{kozlov_eq3.5}
\lambda_2\to -\frac{m\xi u_0}{J_0+m\xi^2}.
\end{equation}
The unbounded decrease in one of the eigenvalues corresponds to the limit
passage to the nonholonomic dynamics of the Chaplygin sleigh (taking into
account the boundary layer phenomenon~\cite{kozlov_10,kozlov_9}). The
limit relation~\eqref{kozlov_eq3.5} gives a~formula for the eigenvalue of
steady-state solutions of the nonholonomic\linebreak
equations~\eqref{kozlov_eq2.5}.

\section{The general case}
\label{kozlov_sec4}

These observations can be generalized. Eqs.~\eqref{kozlov_eq1.4} are a~closed
system of ${n-m}=p\ge2$ differential equations with quadratic right-hand
sides. Indeed, quasi-velocities can be chosen in such a~way that the linear
functions of constraints take the form:
$f_1=\omega_{n-m+1},\ldots,f_m=\omega_n$. Then the first~$n-m$ equations
of~\eqref{kozlov_eq1.4} form a~closed system with the quadratic right-hand side
\begin{equation}
\label{kozlov_eq4.1}
\dot{\omega}=v(\omega),\quad \omega=\left(\omega_1,\ldots,\omega_{n-m}\right)^{\top}\in\mathbb{R}^p,
\end{equation}
and Lagrange multipliers can be found from the remaining $m$ equations.
The
system~\eqref{kozlov_eq4.1} admits the first integral~$H$ as a~positive definite quadratic form.

The problem of finding nonzero steady-state solutions~$\omega_0$ to the
nonlinear system~\eqref{kozlov_eq4.1} is a~nontrivial algebraic problem.
It depends on the structure constants of the algebra~$g$, the inertia
tensor, and the linear functions of\linebreak constraints. The homogeneity of the
system~\eqref{kozlov_eq4.1}
\begin{equation}
\label{kozlov_eq4.2}
v(\lambda\omega)=\lambda^2v(\omega),\quad \lambda\in\mathbb{R},
\end{equation}
implies the following simple property: if~$\omega=\omega_0$ is a~steady-state solution, then
the whole straight line
\begin{equation}
\label{kozlov_eq4.3}
\omega=\mu\omega_0,\quad \mu\in\mathbb{R},
\end{equation}
consists of steady-state solutions. We call the straight line~\eqref{kozlov_eq4.3} a~{\it straight line of
steady motions} (which we abbreviate as {\it SLSM}).

The general number of nonzero nonproportional solutions of the algebraic
system of equations
\begin{equation}
\label{kozlov_eq4.4}
v_1(\omega)=0,\ldots,v_p(\omega)=0
\end{equation}
(including the complex solutions) can be estimated using the Bezout
theorem: if the number of these solutions is finite, it does not exceed~$2^{p-1}$.
Indeed, from the algebraic system~\eqref{kozlov_eq4.4} we select~$p-1$ equations
so that the number of their nonproportional solutions (with the complex
ones) is finite. Since the number of variables equals~$p$ and all
functions $v_1,\ldots,v_p$ are homogeneous in~$\omega$ with homogeneity degree~$2$, then (by the Bezout theorem) the number of
solutions (with multiplicities taken into account) equals~$2^{p-1}$. All these
solutions should satisfy one more (omitted) equation. Therefore, their
general number obviously does not increase.

This estimate is rough and does not take into account the existence of
the positive definite quadratic integral of the system~\eqref{kozlov_eq4.1}.

\begin{theorem}\label{kozlov_theorem1}
If $p=2$, then either all positions are steady {\rm
(}$v_1=v_2\equiv0${\rm )} or~there is only one SLSM.
\end{theorem}

Thus, if $p=2$, the phase portrait of the system~\eqref{kozlov_eq4.1} is
similar in appearance to the phase portraits in
Suslov's and Chaplygin's problems. Theorem~\ref{kozlov_theorem1} shows
universality of stability changes in EPS systems
with two nonholonomic degrees of freedom when the direction of motion is
reversed. Theorem~\ref{kozlov_theorem1} also shows that the general estimate of the
number of SLSMs based on the
Bezout theorem is not accurate.

\proof of Theorem~\ref{kozlov_theorem1}. By a linear change of independent
variables the first integral of the system~\eqref{kozlov_eq4.1} is reduced to the
form
\[
H=\frac{1}{2}\left(\omega^2_1+\omega^2_2\right)\!.
\]
In these variables, let
\[
v_i=a_i\omega^2_1+b_i\omega_1\omega_2+c_i\omega^2_2,\quad i=1,2.
\]
The condition $\dot{H}=0$ gives the following relations for the
coefficients: $a_1=c_2=0$, $b_1=-a_2$, $c_1=-b_2$. Hence, Eqs.~\eqref{kozlov_eq4.1}
become
\begin{equation}
\label{kozlov_eq4.5}
\dot{\omega}_1=\omega_2(b_1\omega_1+c_1\omega_2),\quad
\dot{\omega}_2=-\omega_1(b_1\omega_1+c_1\omega_2).
\end{equation}
If $b_1=c_1=0$, then $v_1=v_2\equiv0$. But if $b^2_1+c^2_1\ne0$, there is
only one SLSM: $b_1\omega_1+c_1\omega_2=0$. QED.

Eqs.~\eqref{kozlov_eq4.5} can be given the conformally Hamiltonian form
\[
\dot{\omega_1}=\rho\frac{\partial H}{\partial \omega_2},\quad
\dot{\omega_2}=-\rho\frac{\partial H}{\partial \omega_1};\quad
\rho=b_1\omega_1+c_1\omega_2.
\]
They take the canonical form after rescaling time as $d\tau=\rho\,dt$. The
multiplier~$\rho^{-1}$ serves as the density of the integral invariant of the
system~\eqref{kozlov_eq4.1}. However, this function has singularities on the
SLSM and reverses sign when crossing this straight
line. Conformally Hamiltonian systems  naturally arise in problems of
nonholonomic mechanics~\cite{kozlov_8,kozlov_bfm}.

Let $\omega=\omega_0$~be a~steady motion. Set
\begin{equation}
\label{kozlov_eq4.6}
A=\left.\frac{\partial v}{\partial \omega}\right|_{\omega_0}.
\end{equation}
This is the linearization operator of the system~\eqref{kozlov_eq4.1} at point~$\omega_0$. Differentiating the identity~\eqref{kozlov_eq4.2} with respect to~$\lambda$ and
then assuming that $\lambda=1$, we obtain
\[
\frac{\partial v}{\partial \omega}\,\omega=2v(\omega).
\]
If $\omega=\omega_0$, we have $A\omega_0=0$. Therefore, the operator~\eqref{kozlov_eq4.6}
is degenerate. By~$u(\omega_0)$ ($s(\omega_0)$) we denote the number of eigenvalues
of the operator~\eqref{kozlov_eq4.6} lying in the right (respectively, left) complex
half-plane. The following simple theorem holds:

\begin{theorem}\label{kozlov_theorem2}
If $\omega_0\ne0$~is a~steady motion, then for $\mu>0$
\[
u(\mu\omega_0)=u(\omega_0),\quad
s(\mu\omega_0)=s(\omega_0),
\]
and for $\mu<0$
\[
u(\mu\omega_0)=s(\omega_0),\quad
s(\mu\omega_0)=u(\omega_0).
\]
\end{theorem}

Indeed, according to~\eqref{kozlov_eq4.6}, the eigenvalues of the linearization operator of at the stationary point~$\mu\omega_0$ differ from
those of the operator~$A$ by the multiplier~$\mu$.

In the case that is the most important to us
\[
s(\omega_0)=n-m-1.
\]
Then (with the zero root) $u(\omega_0)=0$ and,~by Lyapunov's
theorem, the equilibrium point~$\omega_0$ is stable. Moreover, the perturbed
motion infinitely approaches one of the
equilibria~\eqref{kozlov_eq4.3} as $t\to+\infty$. On the other hand, by
Theorem~\ref{kozlov_theorem2}, $u(-\omega_0)$ also equals $n-m-1>0$. Therefore, the
steady motion $\omega=-\omega_0$ is unstable.

We call the SLSM~\eqref{kozlov_eq4.3} \textit{nondegenerate} if at all points of this
straight line, ${\omega\ne0}$, the operator~\eqref{kozlov_eq4.6} has only one zero
eigenvalue. We define the \textit{index} of the nondegenerate SLSM:
\begin{equation}
\label{kozlov_eq4.7}
i=j(\omega_0)+j(-\omega_0),\quad \omega_0\ne0,
\end{equation}
where $j(\omega_0)=1$ ($=-1$) if $s(\omega_0)$ is even (odd). By
Theorem~\ref{kozlov_theorem2}, the index is independent of the choice of point
$\omega_0\ne0$ on the straight line~\eqref{kozlov_eq4.3}. The index can take the
following values: $-2$, $0$, $2$. If~$p$ is even, then (by
Theorem~\ref{kozlov_theorem2}) $i=0$.

\begin{theorem}\label{kozlov_theorem3}
If all SLSMs are nondegenerate, then the sum of their indices equals $1+(-1)^{p+1}$.
\end{theorem}

In order to prove it, we fix the positive level set of the energy integral ${H=h>0}$ and
restrict the initial dynamical system~\eqref{kozlov_eq4.1} to it.
It is only the equilibrium points of the reduced system on the $(p-1)$-dimensional ellipsoid
$\{H(\omega)=h\}$ that coincide with the points of intersection of the
SLSM~\eqref{kozlov_eq4.3} with this ellipsoid. In view of the assumption about
nondegeneracy of the SLSM, all equilibria of the reduced system are
isolated. The index of the singular point~$\omega_0$ of the system on the energy
ellipsoid is defined as the sign of product of the roots of the characteristic
polynomial of the linearization operator for the reduced system at point~$\omega_0$. It is clear that these numbers coincide with the
eigenvalues of the operator~\eqref{kozlov_eq4.6} other than zero. Their product
sign obviously coincides with the number~$j(\omega_0)$ from~\eqref{kozlov_eq4.7}. It remains to
use the Poincar\'{e}--Hopf theorem: the sum of the indices of isolated
singular points of a~dynamical system on a~closed manifold
is equal to its Eulerian characteristic. Recall that the Eulerian characteristic
of the $(p-1)$-dimensional sphere is equal to zero for even~$p$ and is
equal to~$2$ for odd~$p$. QED.

\begin{corollary}\label{kozlov_cor1}
If $p$ is odd, then there is at least one SLSM.
\end{corollary}

\begin{corollary}\label{kozlov_cor2}
If $p$ is odd and all SLSMs are nondegenerate, then their number is odd. In
particular, it does not exceed $2^{p-1}-1$.
\end{corollary}

Indeed, for odd $p$ the index of the SLSM can be either $-2$ or~$2$. If the
number of nondegenerate SLSMs is even, then the sum of the indices is
divisible
by~$4$. However, (by Theorem~\ref{kozlov_theorem3}) this sum equals~$2$. Further,
by the Bezout theorem, there are at most $2^{p-1}$ nondegenerate
SLSMs. But for $p\ge2$ this is an even number. Hence, the total number of
SLSMs does not exceed $2^{p-1}-1$.

It is not improbable that the conclusions of Corollaries~\ref{kozlov_cor1} and~\ref{kozlov_cor2} hold for
even values of~$p$, too. In any case, this holds for $p=2$
(Theorem~\ref{kozlov_theorem1}). When $p=3$, there is either one or three
nondegenerate SLSMs. The latter case takes place for the Euler top when
the inertia moments are unequal.

In conclusion, we mention that a nonholonomic system of two coupled bodies (called a {\it nonholonomic hinge} and generalizing the Suslov problem) is considered in \cite{Shel}.
Evidently, the above methods can be applied to this problem as well.

\end{document}